Cluster-based Superconducting Tunneling Networks

Yury N. Ovchinnikov [1,2] and Vladimir Z. Kresin[3]

[1]*L. D. Landau Institute for Theoretical Physics, Russian Academy of Sciences, 117334 Moscow, Russia*

[2]*Max-Planck Institute for Physics of Complex Systems, Dresden D-01187, Germany*

[3]*Lawrence Berkeley Laboratory, University of California at Berkeley, Berkeley, California 94720, USA*

Abstract

A 2D tunneling network consisting of nanoclusters placed on a surface is studied. It is shown that such a network is capable of transferring large supercurrent at high temperatures. For a realistic set of parameters the damping is quite small, and the smallness is due to strong renormalization of the capacitance of a cluster. The critical field also turns out to be large.

*Introduction* . This paper is concerned with nanocluster-based superconducting tunneling networks. In our previous paper [1] we described the Josephson tunneling between two nanoclusters. The present paper is a continuation of this study [1]. Specifically, we focus on a two-dimensional tunneling network formed by superconducting clusters. Recent progress in "soft landing", that is, in deposition of metallic nanoclusters on a special substrate without disturbing structure of the former (see, e.g.,[2]) makes the idea of such a network realistic. As was demonstrated in our papers [3,4] and recent publications [5,6], , the presence of electronic energy shells in the nanoclusters leads to the appearance of a high - temperature superconducting state.

The main question that will be addressed here is whether the network is capable of transferring a supercurrent, or this current will be damped out. We consider two major factors which may impact the current's amplitude : first, the statistical distribution of current density, and then the impact of quantum fluctuations . It will be demonstrated that the tunneling network is capable of transferring a supercurrent with high current density and at high temperatures. In addition, we evaluate the impact of an external magnetic field on the network.

*Statistical distribution.* Consider a 2D tunneling network. Let us assume that the distribution of the critical current $j_c$ has the Gaussian form:

$$\tilde{w} = (\pi\delta)^{-1/2} \exp[-(j_c^0 - j_c)^2/\delta)] \qquad (1)$$

Here $j_c^0$ is the average value of the critical current, and $\delta/2 = \langle (j_c^0 - j_c)^2 \rangle$. Therefore, the probability $W(j_c \geq \tilde{j})$ for one pair of clusters to have the critical current larger than some value $\tilde{j}$ is given by the expression:

$$W(j_c \geq \tilde{j}) = (\pi\delta)^{-1/2} \int_{\tilde{j}}^{\infty} d\tilde{I} \exp[-(j_c^0 - \tilde{I})^2/\delta] \qquad (2)$$

Therefore, one can write the following expression for the probability $W_n$ for the chain containing n junctions to have the value of the critical current $j_c \geq \tilde{j}$:

$$W_n(j_c \geq \tilde{j}) = W^n \qquad (3)$$

W is described by Eq.(2). Then one can calculate the distribution function $w_n = -\partial W_N/\partial \tilde{j}$ which has a form:

$$w_n = (n/(\pi\delta)^{1/2}) \exp(-F) \qquad (4)$$

$$F = \{\tau_{\tilde{j}}^2 - (n-1)\ln\left[(\pi\delta)^{-1}\int_{\tilde{j}}^{\infty} dj \exp(-\tau_j^2)\right]\}; \tau_j = (j_c^0 - j)\delta^{-(1/2)}$$

The average value of the critical current for the chain containing n junctions is:

$$<j>_n = \int_0^{\infty} dj\, j\, w_n \qquad (5)$$

where $w_n$ is described by Eq.(4). The value of $<j>_n$ can be calculated by the method of descent. Correspondingly, one can determine $\tau_{j_{extr..}}$ which is the solution of the equation

$$2\kappa = (n-1)\pi^{-1/2}\exp(-\kappa^2)[0.5 + \pi^{-1/2}\int_0^\kappa dy \exp(-y^2)]^{-1} \tag{6}$$

where $\kappa \equiv \tau_{j_{extr..}} = (j_c^0 - j_{extr.})/\delta^{1/2}$. For example, for n=2, one can find from Eqs.(5),(6) that $<j>_3 = j_c^0 - (\delta/2\pi)^{1/2}$, that is, an increase in a number of junctions leads to a decrease in the value of the average critical current. For n>>1 Eq.(6) can be reduced to the form:

$$\kappa \approx \ln^{1/2}(n/2\pi^{1/2}\kappa) \tag{6'}$$

One can see that the dependence on n is described by slow logarithmic law. Superconducting current can persist up to very large value $N_{max.} \approx 2\pi^{1/2}\tau_0\exp(\tau_0^2)$; $\tau_0 = (j_c^0/\delta^{1/2})$. Indeed, $n_{max.}$ is very large even for the broadening $\tau_0^{-1} \approx 0.1$.

*Current through the network.* Quantum fluctuations are a major factor leading to the damping of the Josephson current. We have studied the effect of Coulomb blockade for the case of a single junction [1]. Here we focus on the network containing similar superconducting nanoclusters. As we know, the impact of quantum fluctuations is greatly affected by the value of capacitance. This feature has been studied in [7], and also by Larkin and one of authors in [8]. It turns out, and this is the fundamental quantum feature, especially important for nano-based networks, that it is necessary to take into account the renormalization of the capacitance relative to its intrinsic ("bare") value c. According to [7],[8], the renormalized value C is equal to :

$$C = c + Z_c; \quad Z_c = \frac{3e\hbar j_c}{16\varepsilon_0^2} \qquad (7)$$

where c, $j_c$, and $\varepsilon_0$ are the capacitance, the density of the current and the energy pairing gap for an isolated junction

(see below), respectively.

The current $j_c$ was evaluated in our previous paper [1]. For the "magic" ( or near "magic") cluster its geometry is close to being spherical. Then the expression for the current has a form [1]:

$$j_c = \frac{e\hbar^3}{2m^2} T \sum_{\nu\nu_1} \sum_{\omega_n} |T_{\nu,\nu_1}|^2 \frac{|\Delta^L||\Delta^R|}{[\omega_n^2 + (\varepsilon_\nu^L)^2][\omega_n^2 + (\varepsilon_{\nu_1}^R)^2]} \qquad (8)$$

Here $\omega_n = (2n+1)\pi T$ (we employ the thermodynamic Green's functions formalism, see, e.g.,[9]), $\Delta = \Delta(\omega_n)$ is the pairing order parameter, and $\varepsilon_p^i = [(\xi_p^i)^2 + |\Delta|^2]^{1/2}, \xi_p^i = E_p^i - \mu$ is the electronic energy (in the absence of pairing; $i = \{L,R\}, p = \{\nu,\nu_1\}$ ) referred to the chemical potential $\mu$, $\nu, \nu_1$ are the quantum numbers (see Table I), $T_{\nu\nu_1}$ is the tunneling matrix element which has a form (see, e.g., [1]):

$$|T_{\nu,\nu_1}|^2 = |\int_S d\vec{S}[f_{\nu_1}^*(\partial f_\nu/\partial \vec{r}) - f_\nu(\partial f_{\nu_1}^*/\partial \vec{r})]|^2 (\int d\vec{r}|f_\nu|^2)^{-1}(\int d\vec{r}|f_{\nu_1}|^2)^{-1} \qquad (9)$$

$f_\nu$ and $f_{\nu_1}^*$ are the eigenfunctions of the Hamiltonian

$\hat{H} = -(\hbar^2/2m)\partial^2/\partial r^2 + V_i(r) - \mu$ for the left and right electrodes (clusters), respectively (see [1]). For r>a they have a form :

$$f_\nu = \eta Y_l^m K_{l+1/2}(pr)/(pr)^{1/2} \quad (r > a) \tag{10}$$

Here $Y_l^m, J_{l+1/2}$, and $K_{l+1/2}$ are spherical and Bessel functions, $p = (2m\delta U_o)^{1/2}, \delta U_o = \delta U - E_H$, $E_H$ = $E_{HOS}$ is the energy of highest occupied shell, $\delta U$ is the height of the barrier, and *a* is the cluster radius. Expression (10) can be used also as a first approximation for slightly deformed clusters. The constant $\eta$ can be determined with use of the usual boundary conditions at *r=a* and is equal to

$$\eta = -(E_H/\delta U_o)J_{l-1/2}(\kappa a)[K_{l-1/2}(pa) + (l+1)K_{l+1/2}(pa)/(pa)]^{-1}$$

(11)

$\kappa = (2mE_H)^{1/2}$, the notations see in Table I. With use of Eqs.(8)-(11), we obtain the following expression for the critical current:

$$j_c = \frac{e\hbar^3}{m^2} \cdot \frac{E_H}{\delta U_o}(p^2 a^6)^{-1} T \sum_{\omega_n} \sum_{L,L_1} |\Delta(\omega_n)|^2 R_{\omega_n;L} R_{\omega_n;L_1}$$

(12)

$$R_{\omega_n;L} = (2l+1)z_0(l)K_{l+1/2}^2(pd/2)(\omega_n^2+\varepsilon_L^2)^{-1}[K_{l-1/2}(pa)+(l+1)(pa)^{-1}K_{l+1/2}(pa)]^{-2} \quad (12')$$

d is the distance between neighboring clusters, $z_0$ are zeros of the Bessel function.

Note, that the discrete nature of the spectrum leads to possibility of resonant tunneling; this special case was also studied in [1]. However, here we focus on the more general case of non-resonant tunneling.

To study the issue of damping, we employ the method developed by Larkin, Schmid and one of the authors [10]. They considered a 2D ordered network of Josephson junctions. The damping is caused by motion of defects (vortices), and in the quantum picture such motion corresponds to barrier tunneling. Then the problem is reduced to the calculation of the effective action, since the damping $\Gamma \propto \exp(-S)$, and the action S is determined by the

expression [ 10 ]:

$$S = 4.5[(\hbar^2/8e^3) j_c Z_c]^{1/2} \qquad (13)$$

where $j_c$ is the current amplitude determined by Eq.( 12), and $Z_c$ is the change (renormalization) in the capacitance (Eq.(7)).

The modified (renormalized) capacitance and, consequently the action depends also on the value of the pairing energy gap (see Eqs.(7), (13)) .The evaluation of this parameter was described by us in [3,4].The pairing energy gap is determined as the root of the equation $\varepsilon = \Delta(-i\varepsilon)$, $\Delta(\omega)$ is the order parameter: $\Delta(\omega) = B\tilde{\Omega}\,[1 + D(\omega/\tilde{\Omega})^2]^{-1}$. The constants B and D could be calculated for any cluster (see [ 4 ]) .

Based on Eqs. (7),(12),(13), one can calculate the action for specific cluster-based network. Consider, for example, the network containing the clusters with the following realistic set of parameters (see table I; this specific case was described in our papers[1],[4]):

$$l_H = 7, l_L = 4, \text{N=168}, a \cong 6\dot{A}, \tilde{\Omega} \cong 25 meV, B \cong 0.45; D \cong 6\bullet 10^{-2};$$

these parameters are also close to those for $Al_{56}$ cluster. We assume also that $\delta U_0$=0.75eV, d= 15A. The straightforward calculation, based on Eqs,(7),(12),(13) leads to the following value for the action: $S \cong 10$. If $\delta U_0 \cong 1eV$, we obtain $S \cong 5$. The general functional dependence $\Gamma(\delta U_0)$ will be described elsewhere.

One can see that S>>1; the damping $\Gamma \sim \exp(-S)$ is small. Therefore, the network can transfer rather large current without any noticeable damping. For example, for the case considered above, the current density $j_c \approx 10^9$ amp/sm$^2$.

Note that we consider here the non-resonant channel. If it possible to build the network transferring the current through the resonant channel (see [1]), then the value of the current could be increased by $\sim 10^2$-$10^3$.

*2D cluster-based network in an external magnetic field.*

Consider the in-plane cluster-based tunneling network in an external magnetic field which is perpendicular to the plane. The problem is similar to that described in [11] (see also [12]). Let us start with the Maxwell equation, which can be written in the form [12]:

$$\Delta \vec{A} = -(4\pi/c\hbar) j_c \delta(z)[r^{-1} - (2eA/c)]\vec{e}_\theta \qquad (14)$$

Here $\vec{A}$ is the vector potential (the gauge is

$$div\vec{A} = 0); \vec{e}_\theta = (-\sin\varphi, \cos\varphi) \qquad (14')$$

Using the Fourier transformation we obtain, after some calculation, the following expression for the current density:

$$\vec{j} = (\lambda_{eff} j_c / r) \int_0^\infty dq J_\circ(qr)(1+\lambda_{eff} q)^{-2} \vec{e}_\theta \quad (15)$$

Here $J_0(x)$ is the Bessel function, $\lambda_{eff}^{-1} = 4\pi e j_c / \hbar c^2$

Eq.( 14 ) allows to obtain the following expressions

(cf.[ 12 ]):

$$\vec{j}(r << \lambda_{eff}) = (j_c / r)\vec{e}_\theta; \vec{j}(r >> \lambda_{eff}) = (\lambda_{eff} j_c / r^2)\vec{e}_\theta \quad (16)$$

For example, for the specific case with d=14 , $\delta U_0$ =1eV we obtain $\lambda_{eff.} \approx 7 \cdot 10^{-4} cm$ . Then one can determine the value of the characteristic magnetic field $H_1$ which corresponds to the overlap of single vortices. It is determined by the relation

$H_1 = \Phi_0 / \lambda_{eff}^2$ ($\Phi_0$ is the flux quantum) and is rather small: $H_1 \approx 0.4$ G. The most important quantity is the critical field which is defined by the relation: $H_2 = \Phi_0 / d^2$ and corresponds to the pinning phenomena. Unlike $H_1$, the value of $H_2$ is very strong: $H_2 \approx 5.10^2$ T. As a whole, because of such a broad interval between $H_1$ and $H_2$, one should expect a rather weak dependence of the current on magnetic field.

*Discussion.* The analysis carried out by the authors in [1] and in the present paper demonstrates that supercurrent can be transferred through a tunneling network formed out of superconducting nanoclusters. Such transfer implies that the clusters are organized on a surface. Since the presence of shell structure in the electronic energy spectrum is a key factor for the pairing, it is important for the surface – cluster interaction do not perturb the cluster's geometry and, correspondingly, its energy spectrum. This is a serious and well–known challenge (the so called "soft landing" problem), but recent progress with the use of, for example, $C_{60}$–based substrates [2] makes it realistic to envision such tunneling networks.

One should mention also the possibility of building 3D network ; such an idea was proposed in [13]. This picture is based on 3D crystal with Josephson current flowing between the cluster units. A possible example of such a system is the crystal formed from ligand-stabilized $Ga_{84}$ clusters [14]. These are different from the systems analyzed in [3,4] which are capable to uphold pairing up to high $T_c$. The crystal studied in [14] displayed $T_c \cong 8K$, which still much higher than that for bulk $Ga (\cong 1.1K)$. The authors suggested that this was due to the mechanism [13].

In summary, development of cluster–based network described in [1] and in present paper is an interesting and promising direction. Using these, one will be able to observe macroscopic supercurrents with large current densities and at high temperatures.

## ACKNOWLEDGMENTS

. The research of Y.O. is supported by EOARD, contract #097006. The research of V.K. is supported by US AFOSR.

Table I (list of notations).

| | |
|---|---|
| $j_c$ | critical current |
| c;C | capacitance for an isolated cluster (c) and its renormalized value (C) for the network |
| L;R | left (L) and right (R) electrodes (clusters) |
| $\nu; \nu_1$ | quantum numbers for the left ($\nu \equiv \nu_L$) and right ($\nu_1 \equiv \nu_R$) clusters |
| $\delta U_0$ | height of the barrier |
| $E_H$ | energy of highest occupied shell ($E_H \equiv E_{HOS}$) |
| $l$ | orbital momentum |
| a;d | cluster radius (a) and the distance between the centers of neighboring clusters (d) |
| S | action |